\documentclass[aps,prb,amsmath,amssymb,showpacs,twocolumn]{revtex4}
\usepackage{graphicx} 
\usepackage{dcolumn} 
\usepackage{array} 
\usepackage{bm} 
\usepackage{latexsym} 

\begin{document} 



\title{Coupling of orthogonal diffusion modes in two-dimensional nonhomogeneous systems}

\author{F Krzy\.zewski, Magdalena A. Za{\l}uska--Kotur}
\email{zalum@ifpan.edu.pl} 
\affiliation{Institute of Physics, Polish
Academy of Sciences, Al.~Lotnik{\'o}w 32/46, 02--668 Warsaw, Poland}

\date{\today}

\begin{abstract}
Collective diffusion coefficient in a two-dimensional lattice gas on a nonhomogeneous substrate is investigated using variational approach. Particles reside at adsorption sites with  different well depths potentials and jump randomly between them.  The site blocking is the only particle--particle interaction mechanism. It is shown that the value of the diffusion coefficient in one lattice direction depends nontrivially  on the rate and the character of the particle jumps in all directions. When the jump rate in the direction perpendicular to that along which the diffusion is observed increases, the collective diffusion coefficient approaches values predicted within the mean field approximation. Results of the Monte Carlo simulations for selected systems are very well reproduced by our analytical results.  
\end{abstract}



\maketitle

\section{\label{sec:A}Introduction}

Collective or chemical diffusion of adsorbed species describes diffusion of the local density of many particle system involving individual jumps from one binding site to another. Diffusion is an important process that controls many physical phenomena  as building of nanostructures, crystal growth or rate of chemical reactions. Analysis of such processes quite often assumes collective diffusion as if particles were independent of each other. Such an approach, although reasonable  as a first approximation, can give quite misleading results if a more precise answer is needed.
From a perspective of a theorist collective diffusion  is a complicated many--body problem of
diffusion to which a variety of approaches
are being applied ranging from analytic ones based on master,
Fokker--Planck, or Kramers equations to numerical Monte Carlo or
molecular dynamics simulations. An important background is provided by
the works of Reed and Ehrlich\cite{reed81}, an early summary by
Gomer\cite{gomer90}, and recent reviews by Danani {\it et
al.}\cite{danani97} and by Ala-Nissila {\it et al.}\cite{alanissila02}.
Relevant analytic results for some generic simple models were collected
by Haus and Kehr\cite{haus87} and interrelations between different
statistical descriptions of these processes have been reviewed by Allnatt
and Lidiard\cite{allnatt87}. 

Our interest is the coverage dependence of the collective diffusion
coefficient in a two-dimensional kinetic lattice gas model. 
Most of the activity in this field has been dedicated to interacting gases on homogeneous substrates with different geometries. 
One of the earliest seems to be a linear response theory approach by
Zwerger\cite{zwerger81} which allowed to derive analytic expressions for the
coverage dependent collective diffusion coefficient, $D(\theta)$, for a 1D lattice gas with NN and NNN interactions. 
 Kreuzer and his collaborators,
using a version of the kinetic lattice gas model which he developed earlier
to study thermal desorption kinetics\cite{kreuzer90a,kreuzer90b},
investigated $D(\theta,T)$ (with $T$ being temperature) in a 1D and 2D lattice gas on nonhomogeneous substrate 
with NN interactions and different models of microscopic
kinetics\cite{payne07,mcewen06}.
 
In this work we discuss 2D systems in which interactions are limited to the site blocking only, but the local site potential energy landscape changes from site to site within one lattice elementary cell, i.e. the underlying substrate is nonhomogeneous. The collective diffusion in such a system is a complicated and difficult for an analytical treatment problem. Mean field treatment of the collective diffusion in a Schwoebel potential is due to Merikoski and Ying \cite{merikoski97a,merikoski97b}. Series of Monte Carlo simulation data have been reported by Masin at. al.   \cite{masin05} and theoretical mean field analysis  of these results based on approach balancing of nonequlibrium particle fluxes were presented by Chvoj at al \cite{chvoj06} 
We have shown recently in Ref. \onlinecite{zaluska07} that the variational approach to collective diffusion, proposed in a series of earlier works\cite{gortel04,zaluska05,badowski05,zaluska05a,zaluska06,yakes07},
provides a very efficient and systematic method of analyzing diffusion in nonhomogeneous 1D systems. 
In this work we extend the results of Ref.\onlinecite{zaluska07} to two-dimensional systems. It appears that the diffusion is not a simple product of  one-dimensional projections in two main directions of the lattice. Whereas the diffusion coefficient of a single particle over such a lattice always factorizes so the diffusion coefficients in both directions can be calculated independently, in a many particle system the site blocking induces dynamical correlations between jumps in different directions. Properly selected variational parameters  allow to obtain the expression for the collective diffusion coefficient which contains all possible jump rates present in the model. When the rate of jumps in one direction increases from zero to infinity, the formula describing diffusion coefficient evolves from the one characteristic for a one-dimensional system  \cite{zaluska07} to the 2D mean field theory result, known from Refs. \onlinecite{merikoski97a,merikoski97b}. 
We show that our variational approach works very well in the described cases by comparing analytical results with the Monte Carlo data.

Correlations of the diffusion in two orthogonal directions appear to be a very important factor responsible to a large degree for a difference between the dynamical properties in narrow channels and bulk materials \cite{narrow,hummer01,berezhkovskii02,chaves-rojo08,liu05,maibaum03}. Transport of molecules though molecular pores is an essential , for its  biological and technological applications, collective process in which
correlations in diffusional modes are responsible for the net rate of molecular transport.  
Results which are presented here allow to explore how such correlations build up.

\section{\label{sec:B}Model}

A system of $N$ particles forming an adsorbate is distributed over a two-dimensional nonhomogeneous substrate. We treat diffusion within the adsorbate using a kinetic lattice gas
model.  Basic assumptions are standard: kinetics of the
microstates of the lattice gas is due to the stochastic hopping of
particles to neighbouring sites, only one particle in the gas hops at
any given instant, an average residence time of particles at the
adsorption sites is much longer than the transit time between the
sites, the transition rates of these hops depend on the potential energy landscape experienced by the hopping particle. Double occupancy is forbidden, particles jump between neighboring sites with transition rates that depend on the initial and final states.
Time evolution of this system is governed by the set of Markovian master rate equations for the probabilities $P(\{c\},t)$ that a microscopic {\it microstate} $\{c\}$ of a lattice gas occurs at time $t$
\begin{eqnarray}
  \label{eq:1}
  &&\!\!\!\!\!\!\!
  \frac{d}{dt} P(\{c\},t) \\ &&\!\!\!\!\!\!\!=
  \sum_{\{c'\}} \left[ W(\{c\},\{c'\})
  P(\{c'\},t) - W(\{c'\},\{c\}) P(\{c\},t) \right]. \nonumber
\end{eqnarray}
$\{c\}$ is understood as a set of variables specifying which
particular sites in the lattice are occupied and which are not.  $W(\{c\},\{c'\})$ is a
transition probability per unit time (transition rate) that the
microstate $\{c'\}$ changes into $\{c\}$ due to a jump of a particle
from an occupied site to an unoccupied neighboring site.
The rates $W$ satisfy the detailed balance conditions: 
\begin{eqnarray}
  \label{eq:11}
  W(\{c\},\{c'\}) \ P^{\rm eq} (\{c'\})= 
W(\{c'\},\{c\}) \ P^{\rm eq }(\{c\}).
\end{eqnarray}
Here, $ P^{\rm eq}(\{c\})$ is the equilibrium probability of a
configuration $\{c\}$. In the absence of
interparticle interactions the rate depends only on the local
potential energy landscape experienced by the hopping particle.
For thermally activated jumps it depends on the difference
between the potential energy of the particle at the top of the
potential energy barrier between the sites involved and that at the
initial site. 

\begin{figure}
\includegraphics[height=5cm]{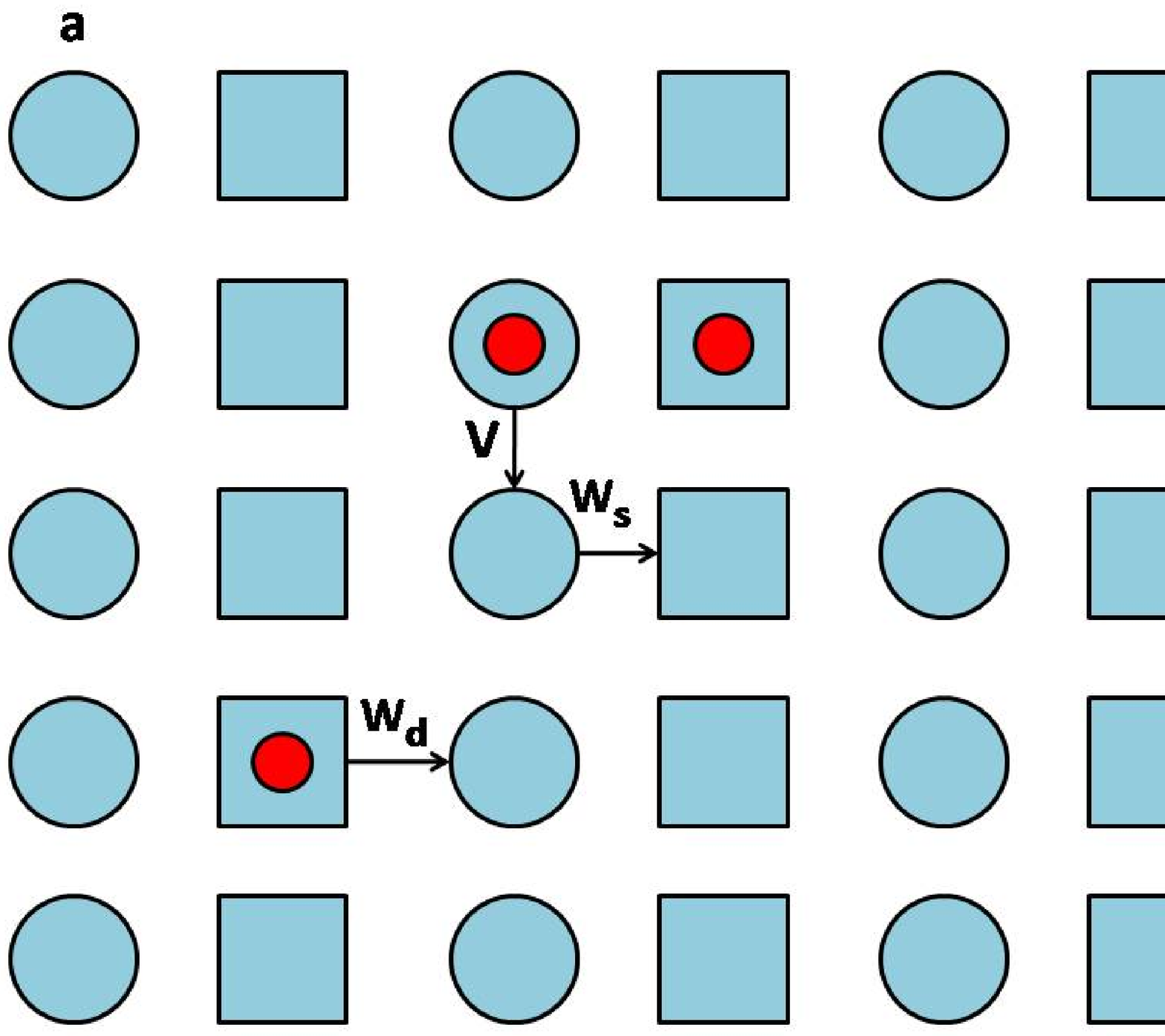}
\hspace{0.5cm}
\includegraphics[height=5cm]{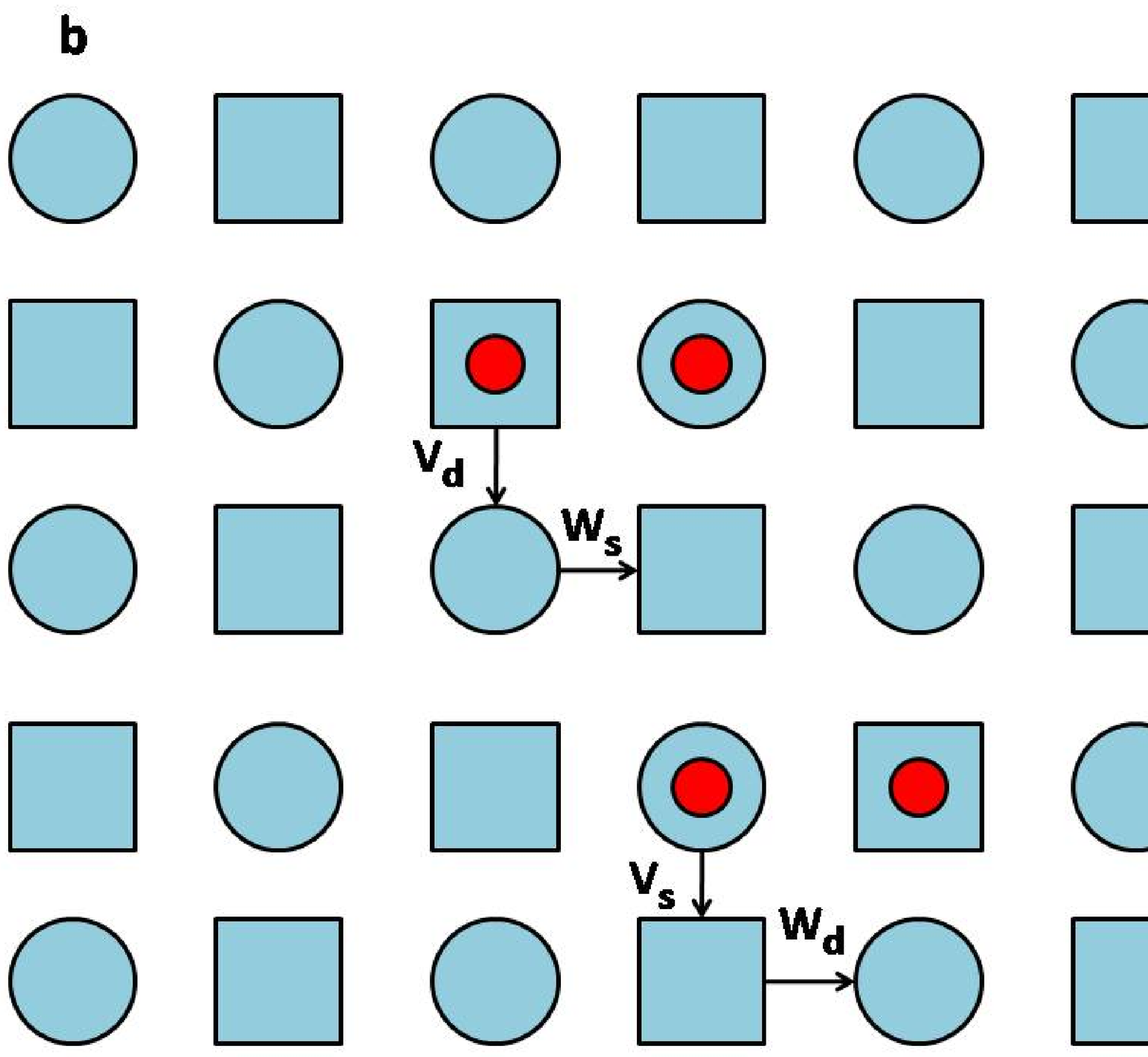}
\caption{Examples of potential geometry of the studied systems a) striped lattice , b) checkered lattice.}
\end{figure}

In order to investigate how the collective diffusion coefficient in a given direction is controlled by the particle jumps  and the geometry of the lattice in the direction perpendicular to it we analyze diffusion over two types of two-dimensional lattices, shown in Fig. 1. 
They consist of periodically repeated patterns of site potentials and intersite barriers. Elementary cell of the striped lattice (Fig 1a) is $2 \times 1$: there are two nonequivalent sites along $x$--axis, and all sites are identical along $y$--axis. Checkered lattice (Fig.1b) has larger, $2 \times 2$ elementary cell with two pairs of identical sites.
In general we have $n_x \times n_y$ unit cell and the potential at each site in the unit cell can be different.  Potential energy  at the tops of all barriers between the sites are assumed to be the same.
We assume that system under study is a box of $L_x \times L_y$ unit cells, i.e. it contains $L_x n_x \times L_y n_y $ sites. We employ periodic boundary conditions and eventually take a $L_x, L_y \rightarrow \infty $ limit.

We employ the variational approach to extract the diffusion coefficient from Eqs (\ref{eq:1}) \cite{zaluska07,gortel04,zaluska05,badowski05,zaluska05a,zaluska06}. To this end, microscopic  states of the systems need to be properly parameterized. 
Following Ref.~\onlinecite{gortel04} we identify a microstate $\{c\}$
by selecting one particle as a {\it reference particle} and specify
positions of all remaining $N-1$ particles with respect to it. Position $\vec X +a \vec l_0$ of the reference particle in systems with nonequivalent sites is specified using two variables \cite{zaluska07}: (i) position $\vec X$ of the elementary cell in which it resides and (ii)  position $a\vec l_0$ within the cell. For two dimensional systems $\vec X$ and $\vec l_0$ are vectors with  two components, $a$ is a distance between the adsorption sites referred to as a fractional lattice constant in what follows.  Then a microstate $\{c\}$ may be identified by the following set of $2(N+1)$ numbers
\begin{eqnarray}
  \label{eq:4}
   \{c\} = [\vec{X}; \vec{\ell}_0: \vec{m}_1, \vec{m}_2, \dots , \vec{m}_{N-1}] \equiv [\vec{X};\{\vec{m}\}],
\end{eqnarray}
where $\vec{m}_j=(m_x,m_y)_j$ is a pair of integers indicating how far, in units of the fractional lattice constant $a$, the $j$-th particle
($j=1,2,\ldots,N-1$) is away from the reference particle. The set of $2N$ integers, $\{\vec{m}\}=[\vec{\ell}_0:
\vec m_1,\vec m_2,\ldots,\vec m_{N-1}]$, is referred to as a {\it configuration}, which  accounts for the relative arrangement of particles in a given
microstate $\{c\}$. The transition rate between two microstates  depends on their configurations only i.e.   
$W(\{c\},\{c'\}) \equiv W_{\{\vec m\},\{\vec m'\}}$. This allows to take an
advantage of the lattice periodicity by taking a lattice Fourier
transform
\begin{eqnarray}
  \label{eq:7}
   P_{\{m\}}(\vec k,t) &=& \sum_{\vec X} e^{i\vec k \vec X} P_{\{\vec m\}}(\vec X,t) 
\end{eqnarray}
of both sides of the rate equations (\ref{eq:1}).  $P_{\{\vec m\}}(\vec X,t)$
stands here for $P\left(\{c\}=[\vec X;\{\vec m\}],t\right)$. It is convenient to
treat $P_{\{\vec m\}}(\vec k,t)$ as an $\{\vec m\}$--th component of a one-column array ${\bf P}(\vec k,t)$ with a macroscopic number of components -- each
component corresponds to an admissible microscopic configuration of
the system. The Fourier--transformed rate equations can be written in
a compact form
\begin{eqnarray}
  \label{eq:8}
   \frac{d}{dt} {\bf P}(\vec k,t) = {\mathbb M}(\vec k) \cdot {\bf P}(\vec k,t),
\end{eqnarray}
where ``$\cdot$'' denotes multiplication following usual ``rows times
columns'' multiplication rules. The matrix elements of ${\mathbb M}(\vec k)$ (referred to as the rate matrix from now on) are, in general, expressed in terms of the rates $W_{\{\vec m\},\{\vec m'\}}$, except for $\{m\}$ to $\{m'\}$ transitions involving a jump of the reference atom across a boundary between neighboring elementary cells. For such jumps elements of $\mathbb M$ are multiplied by an additional $\vec k$ dependent factor $\exp(\pm k_x n_x)$ or $\exp(\pm k_y n_y)$.
For details of the matrix $\mathbb M$ properties and the derivation of all formulas see Refs. \onlinecite{gortel04,zaluska05,badowski05,zaluska05a,zaluska06,zaluska07}.

Eigenvalues of the rate matrix which are always negative account for the temporal decay of a
$\vec k$--th Fourier--component of a density fluctuation from equilibrium.
The one vanishing like $|\vec k|^2$ in the long wavelength limit,
$-\lambda_D(\vec k)$, is referred to as diffusive eigenvalue and yields
the collective diffusion coefficient.  The corresponding eigenvector of ${\mathbb M}(\vec k)$
is referred to as the diffusive eigenvector. 
This eigenvector will be calculated on using variational formula \cite{zaluska07}
\begin{eqnarray}
  \label{eq:17}
  \lambda_D^{\rm var}(\vec k) \equiv \frac{{\tilde {\bm \phi}} \cdot [-{\mathbb
      M}(\vec k)] \cdot {\bm 
      \phi}}
  {{\tilde {\bm \phi}} \cdot  {\bm \phi}} \ge \lambda_D(\vec k)=-D |\vec k |^2,
\end{eqnarray}
where ${\tilde {\bm
    \phi}}$ is trial left eigenvector 
(possibly $\vec k$--dependent) and ${\bm \phi}$ is its right eigenvector
counterpart  with components 
\begin{eqnarray}
  \label{eq:16}
  \phi_{\{\vec m\}} =  P_{\{\vec m\}}^{\rm eq} {\tilde \phi}_{\{\vec m\}}^*.
\end{eqnarray}

For a non--homogeneous substrate, we propose following Ref. \onlinecite{zaluska07} that
the trial left eigenvector has $\{\vec m\}$--th component equal to a sum of phase factors associated with all
{\it occupied sites} in the configuration $\{\vec m\}$:
\begin{eqnarray}
  \label{eq:14}
  {\tilde \phi}_{\{\vec m\}}(\vec k) = e^{i \vec k a(\vec \delta_{\vec \ell_0} + \vec \Delta_{\vec \ell_0})}
    + \sum_{j=1}^{N-1} 
    e^{i \vec k  a(\vec m_j+ \vec \delta_{\vec \ell_j}+\vec \Delta_{\vec \ell_j})}.
\end{eqnarray}
The phase contributed by the $j$--th
particle is determined not only by its distance $a\vec m_j$ from the
reference particle (it is a sole contribution to the phase for a homogeneous system). It receives two additional distinct contributions
$\vec \delta_{\vec \ell_j}=({\delta_x}_{\vec \ell_j},{\delta_y}_{\vec \ell_j}) $ and $\vec \Delta_{\vec \ell_j}                =({\Delta_x}_{\vec \ell_j},{\Delta_y}_{\vec \ell_j}) $ which play a role of the
variational parameters allowing to minimize $\lambda_D^{\rm var}(\vec k)$.
Both depend on the position $a \vec \ell_j$ within an elementary cell of
the site at which the $j$--th particle resides.  The
first one, $\delta_{\vec \ell_j}$, called the {\it geometrical phase}, accounts for a nonhomogenity at the substrate within a unit cell and does
not depend on the presence of other particles in the system. It is always possible to select one particular site (the same within each unit cell) which, if occupied, contributes the geometrical phase $\delta_0=0$.
The other phase, $\Delta_{\vec \ell_j}$, called the {\it
correlational phase}, is introduced to account for correlations between the
$j$--th particle and all the remaining ones and, in principle, it
depends on the state of occupation of all sites in the system. Following previous work \cite{zaluska07}   we assume that it is sensitive to
the occupation of sites nearest to ${\vec \ell_j}$ only.  Thus correlational phases $\Delta_{\vec \ell_j}$, associated with a pair of particles at $\vec{\ell}_{xj}-1,\vec{\ell}_{yj}$ and $\vec{\ell}_j=\vec{\ell}_{xj},\vec{\ell}_{yj }$ will appear in all equations as a sum of contributions to the phase of the particle at the position $\vec \ell_j$ due to its left neighbor and of the phase of the particle  at the position $ \ell_{x j}-1,\ell_{y j}$ due to its right neighbor $\vec \Delta_{\vec \ell_{xj}} ={\vec \Delta_{ \ell_{x j}-1,\ell_{y j}}}^R+{\vec \Delta_{ \ell_{x j} \ell_{y j}}^L}$. Similarly the phase $\vec \Delta_{\vec \ell_{y j}} $ appears with neighbors along direction $y$. Different phases associated with particle pairs are additive.
Summarizing: a phase related to each particle in the
system depends on (i) the distance of the particles from the reference
particle, (ii) an address within the elementary cell of the site which
it occupies (geometrical phase), and (iii) state of occupation of the sites adjacent to it (correlational phase).

  Periodic boundary conditions in two dimensions  imply in the wave
number ($\vec k$) domain that the conditions
\begin{eqnarray}
  \label{eq:41}
  e^{ik_x a n_x L_x}= 1, e^{ik_y a n_y L_y}= 1,
\end{eqnarray}
must be used in the calculations before the long wavelength limit
$|{\vec k}{ a}|^2 \ll 1$ is applied. ($L_x$ - number of unit cells along $x$ direction, $n_x$- number of sites within the cell along $x$).

We see from Eq.~(\ref{eq:17}) that the diffusion coefficient $D_{var}$ is a ratio
\begin{eqnarray}
  \label{eq:20}
  D_{var}=-\frac{\lambda_D^{\rm var}}{|\vec k|^2} = \frac{{\cal M}(\vec k)}{{\cal N}(\vec k)|\vec k|^2},
\end{eqnarray}
of the ``expectation value'' numerator 
\begin{eqnarray}
  \label{eq:21}
\nonumber 	
 {\cal M} (\vec k) &=& \sum_{\{m\},\{m'\}}^{\rm no\ rep} P^{\rm eq}_{\{m'\}}
  W_{\{m\},\{m'\}} \\ 
  &\times&
  \left| {\tilde \phi}_{\{m'\}}^*(\vec k) - {\tilde
      \phi}_{\{m\}}^*(\vec k) \right|^2,
\end{eqnarray}
to the ``normalization'' denominator
\begin{eqnarray}
  \label{eq:22}
  {\cal N}(k) = \sum_{\{\vec m\}} P^{\rm eq}_{\{\vec m\}} \left| {\tilde
    \phi}_{\{\vec m\}}(\vec k) \right|^2.
\end{eqnarray}
Eqs.~(\ref{eq:16}) and (\ref{eq:14}) have been used to
get the final expression for the numerator in Eq.~(\ref{eq:21}). Due to the detailed balance condition (\ref{eq:11}) each
term in (\ref{eq:16}) correspons to a pair of configurations
$(\{\vec m\},\{\vec m'\})$ for transitions from $\{\vec m'\}$ to $\{\vec m\}$ and back.
Each such pair should then appear in the sum only once [as indicated by the
comment ``no rep'' above the sum in Eq.~(\ref{eq:21})] in order to
avoid double counting. In fact,  it was shown in Ref. \onlinecite{zaluska07} that
the dependence of the diffusion denominator
${\cal N}(k)$ on variational parameters $\vec \delta_{\ell_j}$ and
$\vec \Delta_{\ell_j}$ can be ignored in the long wavelength limit and that
\begin{eqnarray}
  \label{eq:23a}
  \lim_{k \rightarrow 0} {\cal N}(k) = \left[ N\left(\frac{\partial(\mu/k_B T)}{\partial \ln \theta}\right)_{T}\right]^{-1} \equiv \left<N^2\right> -
\left<N\right>^2. 
\end{eqnarray}
Here $\mu$ is the chemical potential. The diffusion denominator reduces
to the square of the particle number fluctuation in the system, whereas the numerator ${\cal M}(k)$ depends on the details of particle dynamics and on all variational parameters of the model.

\begin{figure}
\includegraphics[angle=-90,width=9cm]{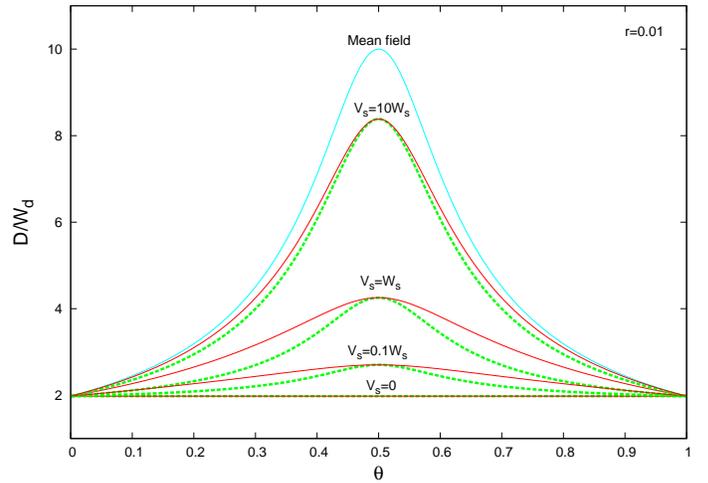} 
\caption{Diffusion coefficient $D_x$ dependence on the total density of the system $\theta$ on striped (dashed line) and checkered lattices (full line). Jump rates along $y$- direction on striped lattices are related to the jump rates on checkered lattices for the corresponding diffusion curves like $V=\sqrt{V_d V_s}=V_s \sqrt{r}$.  
The lowest line corresponds to 1D case, $V=V_s=V_d=0$. The topmost line for $V=V_s=V_d\rightarrow \infty$ reproduces  the mean field result from Refs \onlinecite{merikoski97a,merikoski97b}. }
\end{figure} 
\begin{figure}
\includegraphics[angle=-90,width=9cm]{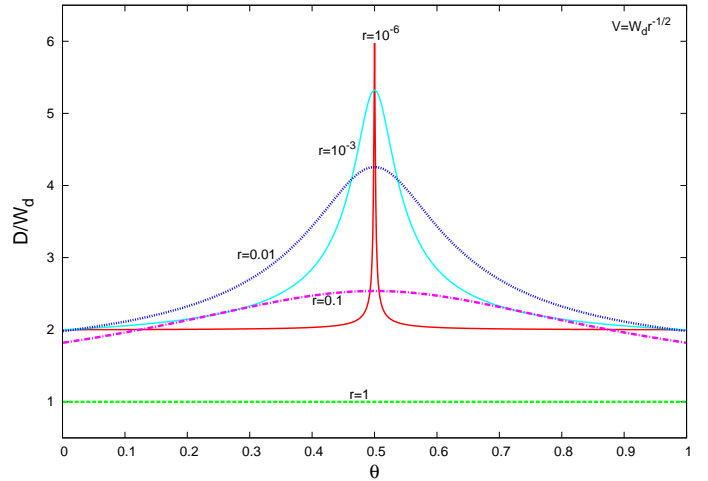} 
\caption{Diffusion coefficient $D_x$ dependence on the total density of the system $\theta$ on striped lattice for different rate $r$ values. The jump rate along $y$-direction is equal to the geometrical mean value of rates in $x$ direction $V=W_d/\sqrt{r}$.  
}
\end{figure}

\section{Striped lattice}

Let us consider first the striped lattice shown in Fig 1a. It consists of rows of sites, with alternating deep and shallow potential energy minima. Transition rates of jumps along $x$--are $W_d$ for a jump from deep well and $W_s$ for a jump from shallow well, whereas all jumps along $y$--axis have  the same rate $V$. A ratio $r=W_s/W_d$ between both rates along $x$--axis is the only parameter  which determines the equilibrium properties of the system at a given density of particles. With the site blocking interactions, preventing double site occupation, the mean equilibrium densities are
\begin{equation}
\label{theta_s}
\theta_s= \frac{r \exp(\beta\mu)}{1+r \exp(\beta \mu)}
\end{equation}
for shallow and  
\begin{equation}
\label{theta_d}
\theta_d= \frac{\exp(\beta\mu)}{1+\exp(\beta\mu)}
\end{equation}
for deep sites. The chemical potential $\mu$ controls the total particle density, understood as a fraction of adsorption sites that are occupied $\theta=(\theta_s+\theta_d)/2$.
The denominator ${\cal N}(0)$ given by Eq (\ref{eq:22}) in the limit ($|\vec k|a)^2\rightarrow 0$ depends only on the equilibrium properties of the system. With no interactions other than the site blocking we have  
\begin{equation}
\label{N}
{\cal N}(0)=\theta_s(1-\theta_s)+\theta_d(1-\theta_d)
\end{equation} 

In order to derive the numerator ${\cal M}(\vec k)$ (\ref{eq:21}) we use  variational vector $\tilde \phi_{\{m\}}(\vec k)$ Eq (\ref{eq:14}). 
There are two geometrical phases: $\delta_d^S=0$ (by choice) and $\delta_s^S$, which for the substrate with potential energies of all barriers being the same is also equal to zero \cite{zaluska07}.
 In the basic cell of striped lattice, there is only one pair of  different sites, which means that there is only one nonzero variational parameter 
$\Delta_{x}^S$. It is the correlational phase of a particle being in a site with a shallow potential well, that has neighbor at left side, in deep well. All other occupational phases are equal to $\pm \Delta_{x}^S $, depending on the order in which particles occupy deep and shallow wells.
Thus components of trial eigenvector are
\begin{equation}
  \tilde{\phi}_{\{m_j\}}(\vec k)=\sum_{j=1}^{N-1} e^{ik_xa(m_{xj}+\Delta_{x }^S)+ik_ya m_{y j}}, 
\end{equation}
with $\Delta_y^S=0, \delta_s^S=\delta_d^S=0 $.
After solving variational equation we get 
\begin{equation}
\Delta_x^S=\frac{W_s-W_d}{W_s+W_d+V}.
\end{equation}

Final expression for the diffusion coefficient along the $x$-direction in the striped system is given by
\begin{equation}
\label{striped_diff}
D_x^S=a^2 \frac{2W_s }{W_s+W_d+4V} \left[ W_d+\frac{4V\theta_s(1-\theta_d)}{\theta_s(1-\theta_s)+\theta_d(1-\theta_d)}\right ]
\end{equation}
while for the $y$-direction it is
\begin{equation}
D_y^S=a^2 V .
\end{equation}

Now, when we set $V=0$ in  Eq. (\ref{striped_diff}), it simplifies  reproducing the result for diffusion in one-dimensional (1D) system \cite{zaluska07}  
\begin{equation} 
D_x^{1D}=a^2 \frac{2W_s W_d}{W_s+W_d}.
\end{equation}
In this limit diffusion does not depend on the density of the system, what is illustrated in the lowest line in Fig. 2. 
For all nonzero values of $V$, diffusion along $x$--axis depends on the total density $\theta$. This dependence changes with the jump rate along vertical direction $V$. Whereas boundary values that represent diffusion of the single particle $\theta=0$ and of the single hole $\theta =1$ stay unchanged irrespectively of the rate in $y$--direction, the height of the diffusion maximum grows up with the increasing value of $V$.  In the limit  $V \rightarrow \infty$ we have the dependence
\begin{equation}
D^{\rm inf}_x=a^2 \frac{2 W_s \theta_s(1-\theta_d)}{\theta_s(1-\theta_s)+\theta_d(1-\theta_d)}. 
\end{equation} 
This relation above reproduces exactly the formula for the diffusion coefficient in the mean field approximation \cite{merikoski97a,merikoski97b}. 

We can see that the rate of the jumps in the vertical direction $V$ controls the character of the density dependence of the diffusion coefficient. With increasing $V$ we observe a smooth transition from purely one-dimensional to the mean field behavior of the system. While the former limit is obvious, the latter one can be understood in such a way, that a particle, capable of fast travel along $y$--axis, detects mean field  occupation value of the neighboring site and instantaneously adjusts to it. 

Character of the density dependence of the diffusion coefficient depends strongly on the ratio $r$ of rates from the deep and shallow sites along the direction under study. In Fig. 3 we show how the diffusion changes with  $r$. The  jump rates along $y$--axis increase with decreasing $r$ like $V=W_d/\sqrt{r}$, slower than the  quicker of two rates along $x$:  $W_s= W_d/r$. We see in Fig. 3 that with such a choice of parameters, the curves become higher and  steeper as $r$ decreases, approaching  to the limiting behavior in which  diffusion coefficient has value $D=2W_d$ for all densities, except  at a one discrete point of $\theta=0.5$ for which $D=6W_d$.

\begin{figure}
\includegraphics[angle=-90,width=9cm]{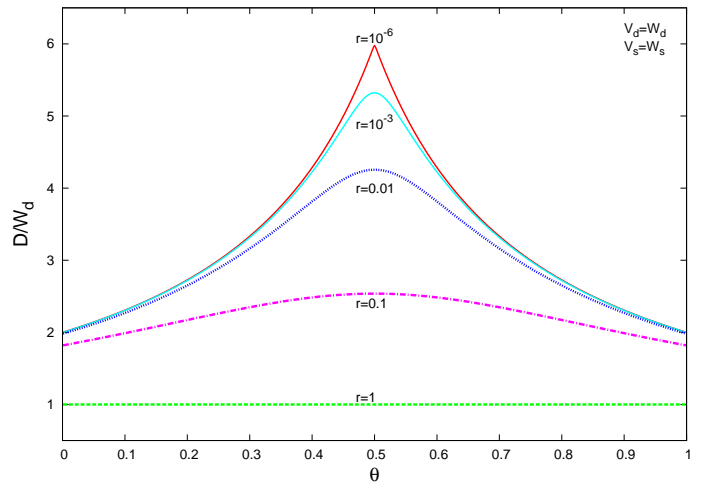} 
\caption{Diffusion coefficient $D_x$ dependence on the total density of the system $\theta$ on checkered lattice for different rate $r$ values. Jump rates along $y$--direction are  equal to the corresponding rates along $x$.  
}
\end{figure}

\section{ Checkered lattice}
Let us now consider a checkered substrate lattice which has the same structure in both $x$-  and $y$--directions. We assume, that like in chessboard every second site is different (see Fig 1b). Jumps out of the shallow sites in $x$--direction are given by $W_s$, and in $y$--directions by $V_s$, whereas jumps out of deep sites are $W_d$ and $V_d$, respectively. The condition 
\begin{equation}
\frac{W_s}{W_d}=\frac{V_s}{V_d}
\end{equation}
has to be fulfilled. The equilibrium occupancies are again given by Eq. (\ref{theta_s}) for the shallow site and by  Eq. (\ref{theta_d}) for the deep site. 
Consequently, the diffusion denominator is again given by Eq. (\ref{N}).
 As before, all geometrical phases $\delta$ are equal to zero due to equal height of all potential energy barriers.
There are now two nonzero, correlational phases. Minimizing the diffusive eigenvalue results in the phase associated with a pair along $x$--direction
 \begin{eqnarray}
&& \Delta_x^C = \\
&&\frac{(W_s-W_d)\theta_s(1-\theta_d)}{ (W_s+W_d)\theta_s(1-\theta_d)+2V_d[\theta_d(1-\theta_d)+\theta_s(1-\theta_s)]}, \nonumber\\
\end{eqnarray}
and the second phase associated with similar  pair of particles  along $y$--direction 
\begin{eqnarray}
&&\Delta_y^C= \\
&&\frac{(V_s-V_d)\theta_s(1-\theta_d)}{ (V_s+V_d)\theta_s(1-\theta_d)+2W_d[\theta_d(1-\theta_d)+\theta_s(1-\theta_s)]}.\nonumber
\end{eqnarray}

The resulting  diffusion coefficient  along $x$--direction is
\begin{eqnarray}
&&D_x^C= \\
&&a^2 \frac{2W_s\theta_s(1-\theta_d)(W_d+2V_d)}{(W_d+W_s)\theta_s(1-\theta_d)+2V_d[\theta_d(1-\theta_d)+\theta_s(1-\theta_s)]}\nonumber
\end{eqnarray}
and for diffusion along $y$ we must replace all $V$ and $W$ rates with $W$ and $V$ respectively.

 $D_x$ as a function of $\theta$  for several values of rates $V_s$ and $V_d$ is plotted in Fig. 2. Maxima of the diffusion coefficient for the striped and checkered lattices are equal if we choose $V= \sqrt{V_s V_d}$. Comparing now curves in both cases, we see that the data for the checkered lattice lie somewhat above the data for the striped lattice, joining together at densities  $\theta=0,0.5$, and $1$. Both models have the same $V \rightarrow 0$ and $V \rightarrow \infty$ limits.

In the Fig. 4 we plot the change in the shape of the density dependence of the diffusion coefficient as a function of the ratio $r$. We keep $V_s=W_s$ and $V_d=W_d$. Now the increase of the diffusion  with decreasing parameter $r$ does not lead to a singular behavior as $r\rightarrow \infty$. Comparing Fig. 3 with Fig. 4 we can see evident qualitative difference in the behavior of both systems. This difference was not so clearly seen in Fig. 2,  where results for different values of the rate $V$ were plotted. 

\begin{figure}
\includegraphics[angle=-90,width=8cm]{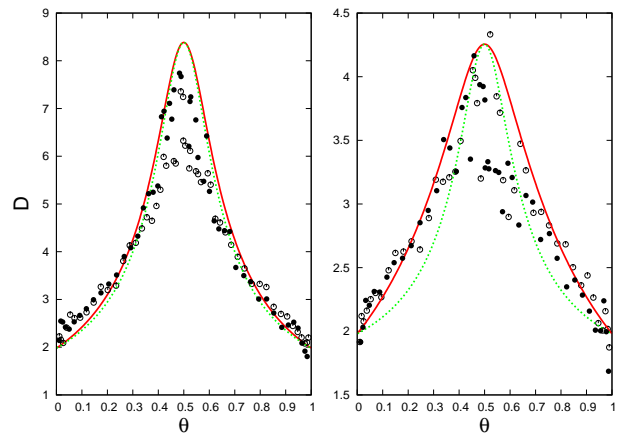} 
\caption{Comparison of Monte Carlo data with analytical curves for the diffusion coefficient $D_x$ dependence on the total density of the system $\theta$. Left panel shows the results for  $V_s= 10W_s$ and right panel for $V_s= W_s$.  Simulation data for striped system are plotted by full circles and analytical curves by dashed line, whereas  simulations for checkered lattices are plotted by open circles and analytical curves by full line. }
\end{figure}

\section{Monte Carlo simulations}
In order to verify  our analytical results for the behavior of the diffusing gas on the nonhomogeneous two-dimensional surfaces we have simulated such systems by using Monte Carlo dynamical approach. We have used
Boltzmann-Matano analysis of the shape of step-like  density profile, after  letting it decay via diffusion process \cite{zaluska99,zaluska00}. Results of this analysis are shown in Fig. 5. Comparison for two different  jump rates along $y$--direction are shown. These rates have been chosen in such a way that $V=\sqrt{V_sV_d}$, so the analytic results merge at $\theta=0.5$. 
In the right panel of Fig. 5 we see system with the same jump rates in both $x$- and $y$--directions. The highest diffusion coefficient is two times larger than  the lowest one. The same difference can be seen for the Monte Carlo data, even if noise of the results is large. If we increase rate of jumps along $y$--direction by a factor of ten, then the expected ratio of the highest and the lowest diffusion coefficient increases to four(left panel). This trend is confirmed by simulations shown in the left panel. Unfortunately due to the high level  of noise the  difference in the behavior between the two types of analyzed lattices  is not clearly visible in the simulation results. However, it is evident that the behavior of the diffusion coefficient as a function of the density changes with the increasing rate of jumps along $y$--axis in both cases in the manner consistent with our analytical results.

\section{Conclusions}
We have shown that the recently formulated variational approach to the collective diffusion is an effective and promising method of calculation of the diffusion coefficient in two-dimensional systems. Here, we have used this method  to describe behavior of a system of particles on a  nonhomogeneous potential landscape. 
The resulting density dependence of the collective diffusion coefficient as a function of all rates that are present in the system is given by  a simple analytic formula. This dependence agrees with Monte Carlo simulation results obtained for selected systems. 
We show that dynamical behavior of two-dimensional system is  interesting and far from trivial even if site blocking is the only interaction that particles experience. 
It appears that  in contrast to a single particle system, collective diffusion in $x$- and $y$--directions of a system of particles depend  on each other in the sense that 
 the diffusion coefficient along one direction strongly depends on the rate jumps of particle in the direction perpendicular to it. This effect is induced by the site blocking, because it is not present in the system of many independently moving particles
We have demonstrated that the one-dimensional character of diffusion changes continuously when the rates of jumps are varied in direction perpendicular to the one along which diffusion is observed. 
We can understand this as a result of an activation of alternative diffusion pathways, when the direct pathway is blocked. 
The transition from the one to two dimensional behavior is highly  nontrivial  even if particles do not interact.

\begin{acknowledgments}
  This work was supported by Poland's Ministry of Science and Higher Education  Grant No.~N202 042 32/1171. The authors would like to thank Dr. Z. W. Gortel for useful discussions and help in preparing this manuscript. 
\end{acknowledgments}

\end{document}